\documentclass[pre,a4paper,twocolumn]{revtex4}
\usepackage[T1]{fontenc}
\usepackage[cp1250]{inputenc}
\usepackage{graphicx}
\usepackage{amsmath}
\usepackage{psfrag}

\newcommand{\bq}{\begin{equation}}
\newcommand{\eq}{\end{equation}}
\newcommand{\ba}{\begin{eqnarray}}
\newcommand{\ea}{\end{eqnarray}}
\newcommand{\dd}{{\rm d}}
\newcommand{\D}{{\rm D}}
\newcommand{\x}{\mathbf{x}}
\newcommand{\G}{\mathbf{G}}
\newcommand{\A}{\mathbf{A}}
\newcommand{\imag}{\mathrm{i}}

\begin{document}
\title{\bf Counting metastable states of Ising spin glasses on arbitrary graphs}
\author{B. Waclaw$^1$}
\author{Z. Burda$^2$}

\affiliation{$^1$Institut f\"ur Theoretische Physik, Universit\"at Leipzig,
Postfach 100\,920, 04009 Leipzig, Germany \\
$^2$Marian Smoluchowski Institute of Physics and Mark Kac Complex Systems Research Centre, \\
Jagellonian University, Reymonta 4, 30-059 Krak\'ow, Poland}

\begin{abstract}
Using a field-theoretical representation of the Tanaka-Edwards integral \cite{tanaka} 
we develop a method to systematically compute the number $N_s$ 
of 1-spin-stable states (local energy minima) of a glassy Ising system with nearest-neighbor interactions and random Gaussian couplings on an arbitrary graph. In particular, we use this method to determine $N_s$ for $K$-regular random graphs and $d$-dimensional regular lattices 
for $d=2,3$. The method works also for other graphs. Excellent accuracy of the results allows us to observe that
the number of local energy minima depends mainly on local properties 
of the graph on which the spin glass is defined.
\end{abstract}

\maketitle

\section{Introduction}
Glassy systems have non-trivial energy landscapes just as many complex
systems observed in nature. The main characteristics of such landscapes 
is the number $N_s$ of local minima, called also metastable states.
Typically, this number grows exponentially with the system size $N$: $N_s \sim e^{Nf_*}$. 
The rate of the exponential growth $f_*$ is a
fundamental quantity characterizing the complexity of the system. It is however very difficult to calculate, 
and it has been analytically found in only a few cases: for
the Ising model with random Gaussian interactions on a complete graph (SK-model) 
\cite{tanaka,bray}, on the one-dimensional closed chain \cite{li,derrida} and for the Ising model 
with random binary interactions $J=\pm 1$ on $K$-regular random graphs \cite{dean}. 

In this paper we describe a method to determine $f_*$ 
for the Ising model with random Gaussian interactions on an arbitrary 
graph. In particular we use this method to compute $f_*$ for some $K$-regular graphs. The idea is to
express the number of metastable states $N_s$ in terms of the Tanaka-Edwards integral 
\cite{tanaka} and then to treat this expression as the partition function of a certain statistical 
field theory. The logarithm of the partition function can be represented as a sum of
connected Feynman diagrams which in turn can be generated and summed on a computer, up to a certain order of the perturbative series.
Using some general properties 
of this series we are able to estimate the value $f_*$ already from
the first few orders with a very good accuracy. 
This allows us to observe that the values of $f_*$ are very similar
for regular graphs with different topologies, indicating that the number of metastable states depends mainly on local properties of the graph.
 
\section{Derivation of the statistical field theory}
We consider a system of Ising spins, $\sigma_i=\pm 1$, $i=1,\ldots, N$, residing on nodes of a simple graph described by an adjacency matrix $\A$. 
The graph does not need to be connected.
The matrix $\A$ is an $N\times N$ symmetric matrix with $A_{ij}=1$ if $i$ and $j$ are connected by an edge or $A_{ij}=0$ otherwise.
The energy of the system is given by:
\bq
	E=-\sum_{i>j} A_{ij} J_{ij} \sigma_i \sigma_j, \label{E}
\eq
where the coupling constants $J_{ij}$'s are random numbers taken from the standardized Gaussian distribution with zero mean and unit variance.

We are interested in counting the number of local minima $N_s$ of the
energy (\ref{E}). 
A local (metastable) minimum is defined as a  configuration of 
spins $\{\sigma_i\}$ such that a flip of any single spin increases energy.
Such a configuration is also called 1-spin-stable.
The number of 1-spin-stable states is given by \cite{tanaka,dean}
\bq
N_s = \sum_{\sigma_1=\pm 1} \cdots \sum_{\sigma_N=\pm 1} 
\prod_{i=1}^N \theta\left( \sigma_i \sum_j A_{ij} J_{ij} \sigma_j \right),
\eq
where $\theta(x)$ is a step (Heaviside) function. 
In Ref. \cite{tanaka} it was shown that the averaging over Gaussian couplings $J_{ij}$
leads to the following concise formula:
\bq
	\left<N_s\right> = \int_{-\infty}^\infty \prod_{k=1}^N \D q_k e^{-\frac{1}{2} \sum_{i,j} M_{ij} q_i q_j}, \label{start}
\eq
where
\bq
	\D q_k = \frac{1}{\pi \imag} \frac{e^{-q_k^2/2}}{q_k-\imag 0^+} \dd q_k,
\eq
and
\bq
	M_{ij} = \frac{A_{ij}}{\sqrt{k_i k_j}} \ .
	\label{MA}
\eq
Here $k_i = \sum_j A_{ij}$ is called degree of node $i$.
The integral (\ref{start}) was calculated  \cite{tanaka} in the limit of $N\to\infty$
for a complete graph: $A_{ij} = 1 - \delta_{ij}$, yielding the result 
\bq
	\ln \left< N_s \right> \cong N f_*,
\eq
with $f_* \approx 0.199228$, known also from earlier considerations
of the SK model \cite{bray}.

In this paper we shall propose a systematic method to evaluate this
integral also for other graphs. Let us introduce an auxiliary
constant $g$ to Eq.~(\ref{start}):
\bq
	\left<N_s\right>(g) = \int_{-\infty}^\infty 
	\prod_{k=1}^N \D q_k e^{-\frac{1}{2} g \sum_{i,j} M_{ij} q_i q_j}, \label{start2}
\eq
The idea is now to find a systematic way of expanding
(\ref{start2}) in powers of $g$ and then to use this series expansion 
to estimate its value for $g=1$:
\ba
	\left<N_s\right> = \left<N_s\right>(g)|_{g=1}.
\ea
Borrowing some techniques from field theory let us define the 
following generating function:
\bq
	Z[J] \equiv \int_{-\infty}^\infty \prod_{k=1}^N \D q_k e^{\imag\sum_k J_k q_k}, \label{zj}
\eq
which allows for rewriting Eq. (\ref{start2}) as:
\bq
	\left<N_s\right>(g) = 
	\left. \exp\left(\frac{1}{2}g \sum_{i,j} M_{ij} \frac{\partial}{\partial J_i} \frac{\partial}{\partial J_j} \right) 
Z[J]\right|_{J=0} .\label{ns2}
\eq
The function $Z[J]$ has a closed form:
\bq
	Z[J] = \prod_{k=1}^N 
	\left[1+{\rm erf}\left(\frac{J_k}{\sqrt{2}}\right)\right] \equiv \prod_{k=1}^N e^{\sum_{n=1}^\infty c_n J_k^n/n!},
	\label{zj2}
\eq
where ${\rm erf}(x)$ is the error function and $c_n$ are cumulants 
of $1+{\rm erf}\left(J_k/\sqrt{2}\right)$. The coefficients $c_n$
can be easily calculated up to an arbitrary order using 
a program for symbolic calculations: $c_1=\sqrt{2/\pi}$, $c_2=-2/\pi$, 
$c_3=2(4-\pi)/(\sqrt{2} \pi^{3/2})$, $...$ . 
Equation (\ref{ns2}) can be graphically represented as a sum 
of vacuum Feynman diagrams of a field theory 
with the propagator $g M_{ij}$ and $\Phi_j^n$-vertices with coupling constants $c_n$. The Feynman rules to calculate the contribution
of a diagram are as follows. To each vertex $\Phi_j^n$, at which $n$ lines meet,
we ascribe a factor $c_n$. The subscript $j$ means that the vertex is decorated
by an index $j=1,\ldots,N$ which can be thought of as a color taken from
a palette of $N$ possible colors. A line joining two vertices decorated with colors $i$ and $j$
contributes a factor $g M_{ij}$. 
Additionally each diagram has a certain symmetry factor which depends on the shape
of the diagram. 
Finally, one needs to perform the summation over colors.

As usual, the logarithm of Eq.~$(\ref{ns2})$,
\bq
	F(g) \equiv \ln \left< N_s \right>(g),
\eq
contains only the contribution from connected diagrams. 
We shall see below that $F(g)$ is an extensive quantity in $N$. 
Thus it is convenient to introduce a density $F_\infty(g)$ per spin which 
for large $N$ becomes a function of $g$ only and can be 
represented as a power series:
\bq
F_\infty(g) \equiv \lim_{N\to\infty} \frac{F(g)}{N} = \sum_{l=1}^\infty f_l g^l. \label{F_g}
\eq
Our goal is to determine $f_* \equiv F_\infty(1)$, which gives 
the rate of exponential growth of the number of 1-spin-stable
configurations for large $N$. 

The coefficients at $g^l$ in the series expansion (\ref{F_g}) come 
from connected Feynman diagrams with $l$ links. 
In the general case, they must be summed on a computer, because their number grows very fast with $l$.
We need a systematic procedure which allows us to sum diagrams in order to calculate the coefficients $f_l$.
In our approach, such a procedure looks as follows:
(A) "draw" all possible connected Feynman diagrams with $l$ links; 
(B) calculate their symmetry factors $s$; 
(C) decorate each vertex of the diagram with an index $i=1,\ldots,N$;  
(D) for each decoration calculate the contribution of the diagram to $F(g)$ as: 
\bq \label{D}
g^l  \cdot s \cdot \prod_{i=1}^v c_{n_i} \cdot \prod_{\left< ij \right>} M_{ij} \ , 
\eq
where $v$ is the number of vertices of the diagram, the first product goes over all vertices and the second one over all links of the diagram; (E) sum contributions of all diagrams and decorations.

Clearly, the procedure described above is not efficient if
the original graph is sparse since then many propagators $g M_{ij}$ are zero. 
Therefore, many of possible $N^v$ decorations give a zero contribution and one wastes time summing many zeros. 
In particular, all decorations of a diagram 
having a self-connecting link give no contribution since $M_{ii}=0$. 
Thus one can omit such diagrams in the sum. One can 
improve the step (C) of the procedure by concentrating only on decorations 
which potentially have a chance to contribute. In other words, one should
look only for decorations for which propagators $M_{ij}$ do not vanish. 
This means that $i$ and $j$ must be neighbors on the graph on which the spins reside. 
Therefore, instead of the step (C) one should take the step (C') in which one only checks such 
decorations which are consistent with the graph structure. This can be done iteratively.
First, one assigns a label $j=1,\ldots, N$ to one vertex of the diagram. 
Then one assigns to its neighbors only values $i$ such that $M_{ij}\ne 0$ or equivalently that $j$ and $i$
are neighbors on the original graph. One repeats this process for neighbors of $i$'s and so on, and selects
only those labels for which the propagator does not vanish. This speeds up the step (E) of the procedure
since now the number of decorations is of order $N\bar{k}^{v-1}$, 
where $\bar{k}$ is the average node degree of the graph.

\begin{figure}
	\psfrag{l=1}{$l=1$}\psfrag{l=2}{$l=2$}\psfrag{l=3}{$l=3$}\psfrag{l=4}{$l=4$}
	\psfrag{s=1/2}{$1/2$}\psfrag{s=1/6}{$1/6$}\psfrag{s=1/8}{$1/8$}\psfrag{s=1/24}{$1/24$}
	\includegraphics[width=6.5cm]{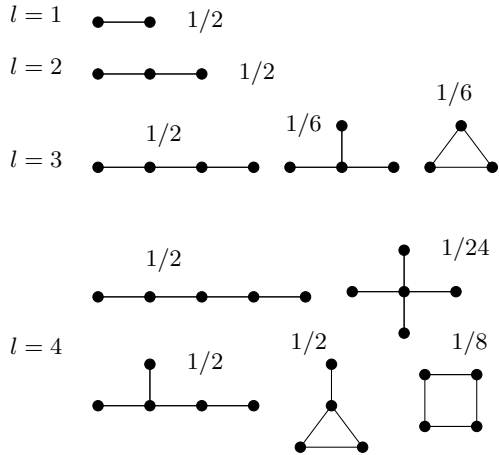}
	\caption{\label{graphs}Feynman diagrams for $l\leq 4$, 
	without multiple links, generated by Eqs. (\ref{ns2}) and (\ref{zj2}), together with their symmetry factors.}
\end{figure}

The procedure (A-C'-E) works for any graph. For a $K$-regular graph
one can essentially simplify calculations since in this case $M_{ij} = A_{ij}/K$, as it stems from Eq.~(\ref{MA}),
and thus the elementary contribution (\ref{D}) to $F(g)$ is  
\bq 
	\left(\frac{g}{K}\right)^l  \cdot s \cdot \prod_i^v c_{n_i} \cdot \prod_{\left< ij \right>}^l A_{ij} \ . 
\eq
The last product of $A_{ij}$'s is either zero or one.
Summing over all decorations of a given diagram we get some number $P$ of decorations consistent with the graph structure.
Each Feynman diagram has now the following contribution:
\bq 
\left(\frac{g}{K}\right)^l  \cdot s \cdot P \cdot \prod_i^v c_{n_i} . \label{DP}
\eq
Actually $P$ is the only part of the expression which depends on the graph structure on which 
spins reside, other quantities can be calculated beforehand. Therefore, we used a C++ program to
generate all simple diagrams (without multiple- and self-connections) up to $l=13$
and applied the routine \texttt{nauty} \cite{nauty} for isomorphism testing to distinguish different
diagrams and to determine their symmetry factors $s$. 
In figure \ref{graphs} we show several examples of diagrams for $l\leq 4$. Their number grows
very fast with $l$: for $l=11$ there are $11461$ such diagrams, for $l=12$ -- $40964$ and for $l=13$ -- $153786$. 
The number of all multi-diagrams, that is diagrams with multiple connections, is much larger but luckily there is no need to generate them.  
Any multi-diagram can be obtained from a simple diagram 
$\gamma$ by replacing its links by multiple links with a certain multiplicity $m$. 
The contribution of all multi-diagrams associated with a simple diagram $\gamma$ is: 
\bq
	s \, P \, \sum_{m_1=1}^{\infty} \cdots \sum_{m_l=1}^{\infty} \left(\frac{g}{K}\right)^{m_1+\dots+m_l}  
	\prod_{a=1}^l  \frac{1}{m_a!} \cdot \prod_{i=1}^v c_{n'_i}
	\label{multic}
\eq
where $s$ and $P$ are calculated for $\gamma$. The terms in the sum contribute to the order
$l'=m_1+\ldots+m_l$ of the expansion of $F(g)$, and $l'$ is the total number of links
of the multi-diagram. Factors $1/m_a!$ are corrections to the symmetry factor which arise from the fact that one can
permute all $m_a$ multiple links joining two vertices without
changing the diagram. $n'_i$ is the number of links meeting at vertex $i$ of the multi-diagram.

\section{A few examples}

\begin{figure}
	\psfrag{1}{$1$}\psfrag{2}{$2$}\psfrag{i}{}\psfrag{j}{}\psfrag{k}{}\psfrag{l}{}\psfrag{m}{}
	\psfrag{a}{a)}\psfrag{b}{b)}
	\includegraphics[width=6.5cm]{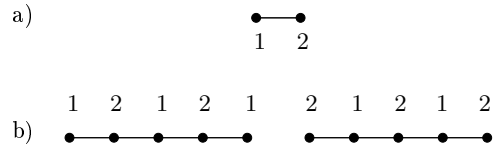}
	\caption{\label{graphs0} (a) A graph consisting of only two nodes $1$ and $2$. (b) The two distinct
assignments of the labels $1$ and $2$ to the linear Feynman diagram for $l=4$. An example of
an inconsistent assignment would be for instance $11212$ since $A_{11}=0$. 
	}
\end{figure}

Let us illustrate how the method works for spins on a graph with $N=2$ vertices. We have $A_{12}=A_{21}=1$, $A_{11}=A_{22}=0$ and $K=1$.
As an example we shall calculate the contribution of the linear diagram with $l=4$ links from Fig. \ref{graphs}. 
Its total contribution is
\bq
	\frac{1}{2} \cdot g^4 \cdot c_2^3 c_1^2 \cdot P, \label{a4gr}
\eq
where the first term accounts for the symmetry factor, 
the second one is a product of four propagators,
the third one is a product of couplings and the fourth one is the combinatorial factor 
$P$ depending on the details of the underlying graph's structure encoded in the adjacency
matrix $\A$. There are only two possible assignments of labels $1$ and $2$ to this diagram, see Fig. \ref{graphs0}. 
All other terms vanish and hence $P=2$.

This example is somewhat artificial because the number of nodes is small. 
Now let us suppose that we have a graph where pairs of nodes $2j$ and $2j+1$ are connected,
so that the only non-vanishing elements of the adjacency matrix are 
$A_{2j,2j+1}=A_{2j+1,2j}=1$ or $j=1,\ldots,N/2$. It is a $1$-regular graph.  
One can easily see that the number of decorations $P$ of any Feynman diagram is $P=N$ or $P=0$.
Indeed, vertices of a given diagram can be alternately decorated with two consecutive numbers
$2j$ and $2j+1$. Thus the choice of a single label specifies automatically the whole decoration.
Because this label may assume $N$ values, we have $P=N$. But if the diagram has 
a loop of odd length, one cannot alternately decorate vertices along this loop, so in this case $P=0$.

Because $P=N$, one can see that also $F(g)$ is proportional to $N$ and hence 
also $\ln \langle N_s\rangle \sim N$. The proportionality coefficient $f_*$ can be determined
in this case analytically by a straightforward calculation of the integral (\ref{start}):
\bq
	F(g) = \frac{N}{2} \ln \left[ \int \frac{dx}{\sqrt{2\pi}} e^{-x^2/2}
	\rm{erfc}^2\left(\sqrt{\frac{g}{2-2g}}x\right)\right],
	\label{fg_1r}
\eq
with ${\rm erfc}(x)=1-{\rm erf}(x)$. We shall use this explicit result to test our method. 
Using the formula (\ref{multic}) and setting $K=1$ we find that the first coefficient in the expansion of
$F(g)$ comes from only one diagram, a line with $l=1$ from Fig.~\ref{graphs}, and reads:
\bq
	s \, P \, c_1^2 = \frac{1}{2} \, N\, \left(\sqrt{2}{\pi}\right)^2 = \frac{N}{\pi} = 0.31831... \cdot N.
\eq
The second coefficient $f_2$ is a sum of the previous diagram with doubled line, and the one for $l=2$:
\bq
	\frac{1}{4} \, N \, \left(\frac{2}{\pi}\right)^2 - 
	\frac{1}{2} \, N \,\left(\sqrt{\frac{2}{\pi}}\right)^2 \frac{2}{\pi} = -\frac{N}{\pi^2} = -0.101321... \cdot N.
\eq
On the other hand, we can calculate $f_l$s numerically from the analytic formula (\ref{fg_1r}):
\ba
	\frac{F(g)}{N} = 
	0.31831 g -0.101321g^2+ 0.096054g^3 \nonumber \\ -0.054306g^4 +0.055831g^5 -0.037248g^6 +... . \label{f6_g}
\ea
We see that both $f_1$ and $f_2$ agree perfectly with those obtained before. 
Higher coefficients $f_l$ can be calculated by performing the sum (\ref{multic}) on a computer. 
We checked that all coefficients, up to $l=11$, agree with those from Eq. (\ref{fg_1r}). 

In the next section we shall make another cross-check by comparing our results 
with those for the celebrated SK model. We shall see that
the method produces correct values of the coefficients $f_l$ as well as of the 
limiting value $f_*=F_\infty(1)$.

\section{The SK model}
The SK model \cite{sk} is the spin glass (\ref{E}) 
on a complete graph, where each node is connected to all other nodes.
Before we apply our procedure of diagrams' summation, let us recall what can be calculated
for the SK model using another methods. The integral (\ref{start2}) can be done 
in the thermodynamic limit \cite{tanaka}, leading to
\bq
	F_\infty(g) = \frac{gt^2}{2} - \ln\left(1+\mbox{erf}(gt/\sqrt{2})\right), \label{fg-compl}
\eq
where $t$ is a solution to a saddle-point equation
\bq
	t\left( 1+\mbox{erf}(gt/\sqrt{2})\right) = \sqrt{\frac{2}{\pi}} \exp\left[ -\left(gt/\sqrt{2}\right)^2\right],
\eq
with $f_*=F_\infty(1)= 0.199228...$. One can find the coefficients $f_l$ by applying Cauchy's differentiation formula and by integrating Eq. (\ref{fg-compl}) numerically. This gives:
\ba
	F_\infty(g) = 0.31831 g - 0.202642 g^2 + 0.147463 g^3 \nonumber \\ - 0.115439 g^4 +  0.094626 g^5 - 0.080058 g^6 + ... \label{f6}
\ea
Let us now calculate $f_l$'s using the method described in Secs. II and III. The propagator $M_{ij}$ for a
complete graph with $N+1$ vertices is $M_{ij} = 1/N$ for any pair of $i\ne j$. It is a $K$-regular graph
with $K=N$, so as one can see from Eq.~(\ref{DP}), each link introduces a suppression factor
$N^{-1}$. On the other hand, the combinatorial factor $P$ contains a power $N^v$, where $v$ is the
number of vertices of the diagram. Thus in the thermodynamic limit, a diagram 
with $l$ links and $v$ vertices gives a contribution $\sim N^{v-l}$. The exponent $v-l$ is equal 
to one minus the number of closed loops in the diagram. 
Therefore, in the limit $N\rightarrow \infty$ only tree diagrams 
give non-vanishing contributions. Our task simplifies therefore to summing 
only tree graphs. Each tree with $l$ links gives the following contribution to $f_l$:
\bq
 s \,g^l \prod_{i=1}^{l+1} c_{n_i} \;,
\eq 
where $s$ is its symmetry factor and $n_i$ is the degree of vertex $i$. 

We performed the summation of all tree diagrams up to $l=11$ 
on a computer and checked that the values $f_l$'s obtained in this way agree 
with those obtained from Eq.~(\ref{fg-compl}). Again, we see that our method gives correct coefficients $f_l$.

We should, however, remember that our goal is not only to find the coefficients of the expansion but
rather $f_*=F_\infty(1)$ which is a sum of infinitely many coefficients. 
If we naively terminate the series at some $L$: $F_L(g) = \sum_{l=1}^L f_l g^l$, 
then e.g. $F_{11}(1)=0.220701...$ is far away from the true value $0.1992...$, because the series is slowly convergent. 
Therefore, we need to find a method which allows to read off the 
limiting value $f_*$ from first few coefficients. As we shall see below one can
find a very good estimate of the limiting value $F_\infty(1)$ using some general information about
the properties of the series expansion. Let us first observe that 
the integral (\ref{start2}) and thus also $F_\infty(g)$
is convergent only if the matrix
\bq
	G_{ij} = \delta_{ij} + g M_{ij} \label{gij}
\eq
is positive definite. In the Appendix A we show that it is so for $|g|<1$ and for $g=1$, and that $\G$ acquires a zero mode for $g=-1$, so the integral (\ref{start2}) is divergent for $g\rightarrow -1$. 
From this we can conclude that the asymptotic behavior of the coefficients 
$f_l$ has the following form:
\bq
	f_l = \frac{(-1)^l a_l}{l^\alpha}, \label{asympt}
\eq
with all $a_l>0$, $0<\alpha\le 1$ and $\lim_{l\to\infty}a_l\to {\rm const}$.
From the asymptotic behavior (\ref{asympt}) we can deduce the following formula for $F_\infty(1)$:
\ba
	F_\infty(1) \cong  F_L(1) + f_L \cdot \left(\frac{L}{2}\right)^\alpha \times \nonumber \\
	\times \left[ \zeta(\alpha,1+L/2)-\zeta(\alpha,1/2+L/2)\right], \label{estim} 
\ea
where $\zeta(\alpha,\beta)=\sum_{k=0}^\infty (k+\beta)^{-\alpha}$ is a
generalized Riemann Zeta function, and $\alpha$ is estimated from the last two coefficients:
\bq
	\alpha \cong  -\left(\ln \frac{-f_L}{f_{L-1}}\right)/\left(\ln \frac{L}{L-1}\right) .
\eq
With the help of formula (\ref{estim}) we can predict now 
the value of $F_\infty(1) \approx 0.199226$ for the SK model. 
To estimate the maximal error we used a method described in Appendix B. 
Our final result $f_*=0.199226(5)$ is in an excellent agreement with the analytic 
result cited above. 

\section{Random $K$-regular graphs and Cayley trees}
A $K$-regular graph is a graph with all degrees equal to $K$. 
We say that a regular graph is random if its adjacency matrix $A$ 
is maximally random under the constraints that $A_{ii}=0,\,A_{ij}=A_{ji}$ 
and $\sum_j A_{ij} = K$ for all $i$. If we fix $K$ and let $N\to\infty$, 
random graphs become sparse and look locally like Cayley trees with degree $K$, 
because the average density of finite-length loops goes to zero in this limit.
Thus for large $N$, instead of computing the coefficients $f_l$ by averaging them over many
$K$-regular random graphs one can calculate them for a single Cayley $K$-tree.  

The propagator $M_{ij}$ is simply $1/K$ if nodes $i,j$ are connected, 
and zero otherwise. Unlike for a complete graph, the contribution 
from diagrams with loops cannot be neglected. Diagrams with multiple 
connections also do not vanish. In order to compute the contribution of a simple 
diagram $\gamma$ and its multi-linked versions we need to evaluate the sum (\ref{multic}).
The summation can be performed on a computer with only a slight complication as compared
to the SK model.

\begin{figure}
	\psfrag{a}{a} \psfrag{b}{b} \psfrag{i}{$i$} \psfrag{j}{$j$}\psfrag{k}{$k$}\psfrag{l}{$l$}
	\includegraphics[width=7.5cm]{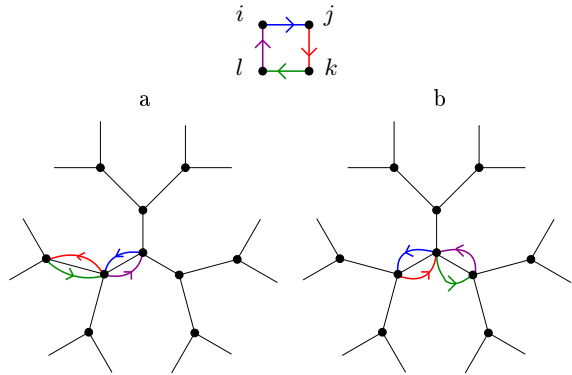}
	\caption{\label{diag_on_graph}
	Example of a Feynman diagram (a square on the top) lied down in two different ways on the Cayley 
	tree with $K=3$. Four links are drawn with four different lines and arrows, 
	in order to show how the diagram is put on the graph. When one vertex of the diagram 
	(say the upper-left) has fixed index $i$, there are only 6 ways of drawing the diagram 
	as in (a), and 9 ways as in (b) case. The picture (a) implies that $j=l$ and there are 
	3 ways of choosing $j$ and 2 of $k$, while for (b) we have $k=i$ and 3 
	possibilities for each index $j,l$. There is no other way of distributing 
	indices $j,k,l$, so $P/N=2\cdot 3+3\cdot 3=15$ for the square diagram on the $3$-tree. }
\end{figure}

The hardest task in the above is to compute $P$. Recall that for a 1-regular graph, $P$ was either $N$ or zero. 
Now $P$ is simply equal to the number of different ways a given Feynman diagram $\gamma$ can 
be lied down on a $K$-tree, in such a way that any two neighboring vertices of the diagram are also neighbors
on the tree. One expects that the number of such possibilities is of order $N K^{v-1}$. The factor
$N$ comes about since one can lie down one vertex of the diagram anywhere on the tree. But the second
has to be located on one of the $K$ neighbors of the first one. So each time by finding a position of
a vertex we should check $K$ neighboring nodes on the tree, which gives the factor $K^{v-1}$.
To illustrate this, consider again the linear graph for $l=4$ from Fig. \ref{graphs}. The diagram 
has $v=5$ vertices. The first vertex can be put anywhere on the tree, but the next one 
only on one of $K$ neighbors of the first vertex. The same for the next one, for which we have again 
$K$ possibilities,  etc., so altogether we have $P=N K^4$.
Let us consider now a more complicated example of a square graph
from Fig. \ref{graphs} and $K=3$. Again we have a trivial factor $N$ for choosing the position 
of the first vertex on the $3$-tree. Once it is chosen, we can position remaining vertices only in $15$ ways as shown in Fig. \ref{diag_on_graph}. 
So we have $P=15 N$ which is less than
$P = 3^4 N$, because of the constraint coming from a closed loop.
Clearly, if a Feynman diagram has no loops then $P=N K^{v-1} = N K^l$. For a diagram with loops
one has to consider constraints on possible decorations.
This can be done by enumerating all possible labellings and accepting only those 
which agree with graph's structure. It can conveniently be done by a computer program. One point must be clarified here - since all $P$'s share the same trivial factor $N$ coming from $N$ possibilities of labeling the initial vertex, 
in the computer program one can just fix one vertex of the diagram to have some arbitrary label, and consider only decorations consistent with this choice.
Next, one multiplies the result by $N$, which then cancels in the definition of $F_\infty(g)$ so that only numbers independent on $N$ remain.

Using this method, we calculated 
$P$'s and then the coefficients $f_l$ up to the given order $L$
(typically $L=11$) for $K=2,3,4$ and $6$. The case $K=1$ has been analyzed in Sec. III.
The results are summarized in Table \ref{results} where we give the values of $f_*$ for $K\geq 2$ 
with estimated errors and compare them with those obtained by numerical simulations based on 
enumeration of all metastable states as described in Ref.~\cite{zbak}. To save space, the coefficients $f_l$
are not shown and can be found elsewhere \cite{www}.

The agreement with simulations is perfect. For the case $K=2$ there is also a beautiful analytic 
result \cite{derrida} to compare with, which gives $f_*=\ln 4/\pi \approx 0.24156...$. As we
see from the table, it agrees with our result within the error bars. We observe also that the 
rate of convergence of the series expansion for $F_\infty(g)$ grows with $K$. 
In other words, for smaller
$K$ one should go to larger order $L$ which is however limited by 
the fast growth of the number of diagrams. This effect is slightly compensated by the fact
that the complexity of computing the combinatorial factor $P \sim K^L$ is smaller for smaller $K$. 

\section{Regular $d$-dimensional lattices}
In this section we shall discuss how to calculate $f_*$ for $d$-dimensional regular lattices, namely
for $d=2$ (square lattice) and $d=3$ (cubic lattice). The lattices are $K$-regular graphs
with $K=2d$, but very special ones. The case $d=1$, that is a closed chain, gives the same 
result as a random $2$-regular graph, because the latter always contains at least one long chain, whose 
contribution to $N_s$ in the limit $N\to\infty$ dominates over the contribution coming from shorter chains. 
In the general case, the only difference as compared to random regular graphs is that now, while calculating $P$,
we shall lie down Feynman diagrams on the $d$-dimensional lattice. Again we see that $P$ is proportional
to $N$ since the first vertex can be put anywhere, but once it is fixed we can put a neighboring
one on a neighboring site of the lattice and repeat it iteratively for all remaining vertices
of the diagram. The calculated values of $f_*$ are given in Table \ref{results}.

\section{Ladders}
As a further example we consider another particular type of $K$-regular graphs:
graphs which we shall call ladders. A ladder is a graph obtained by stacking above each other
infinitely many copies of a $(d-1)$-dimensional cube so that corresponding vertices of the copies
are aligned on a line and the corresponding vertices of consecutive copies are joined by a link. 
A $d=2$ ladder is just what one usually would call a ladder except that it is infinitely long. 
A $d=3$ ladder looks like a bookstand with square-shaped shelves. Such ladders are
$K$-regular graphs with $K=d+1$. It is interesting to compare $f_*=F_\infty(1)$ 
for ladders with those for random graphs and regular $d$-dimensional lattices. 
We adopt the usual scheme. The only thing which changes is again $P$, 
because now we have to lie down diagrams on the ladders. 
The final results are presented in Table \ref{results}.

\begin{table}[ht]
	\begin{tabular}{l|c|c|c|c}
		\hline
		Graph & $K$ & $L$ &\multicolumn{2}{|c}{$f_*$, the slope of $\ln \left<N_s\right>$} \\
		\hline
		& & & result & simulations \\
		\hline
		2-reg. graph & 2 & 13 & 0.2414(2) & 0.242(2) \\
		\hline
		3-reg. graph & 3 & 12 & 0.22484(4) & 0.226(1) \\
		2D ladder & 3 & 12 & 0.22568(6) & 0.226(2) \\
		\hline
		4-reg. graph & 4 & 11 & 0.21762(2) & 0.219(1) \\
		2D lattice & 4 & 11 & 0.21808(2) & 0.219(2) \\		
		3D ladder & 4 & 11 & 0.21799(2) & 0.220(4) \\
		\hline
		6-reg. graph & 6 & 10 & 0.21101(2) & 0.211(1) \\
		3D lattice & 6 & 10 & 0.21125(1) & --- \\
		\hline
		SK model & $\infty$ & 11 & 0.199226(5) & 0.199(1) \\
	\end{tabular}
	\caption{\label{results}Values of $f_*=F(1)$, calculated for various graphs with average connectivity $K$, compared to computer simulations. $L$ stands for the number of coefficients $f_l$ used to estimate $F_\infty(1)$.
All simulations were made for $N=10,\dots,24$, so for relatively small systems. The uncertainty of the last digit is given in brackets $()$. It was estimated as the standard error in case of computer simulations, and as the maximum error in the way presented in Appendix B for semi-analytical calculations.
}
\end{table}

\section{Conclusion}

We presented a semi-analytic method to estimate the rate $f_*$ of exponential growth
of the number of metastable states $N_s$
of the Ising spin glass on different kinds of graphs. 
We checked that the method reproduces results known analytically, 
which are available for a few particular cases.
The method is based on a diagrammatic representation of the quantity $\ln \langle N_s \rangle(g)$
and on the exact enumeration of all Feynman's diagrams up to a given order $L$.
The accuracy of the method improves with growing $L$ but already for $L$ close to $10$ it yields very
precise estimates whose uncertainty varies in the range of order $10^{-4}$ - $10^{-6}$ depending
on $K$ (see Table \ref{results}). The results show that the exponent $f_*$ is determined mainly 
by the degree $K$. This suggests that the number of local minima $N_s$ depends strongly
on local properties of the graph and weakly on its global topology. In other words,
an important information about the complexity of the energy 
landscape of the corresponding spin glass is encoded in the short-range properties of the graph.  
It would be interesting to test if this also holds for other complex systems.

The method presented here can be applied to any type of graphs. However, 
the computational complexity of the method and the dependence of the 
accuracy of $f_*$ on the order $L$ has to be tested case by case.

In this paper we calculated $\langle N_s \rangle$
for Gaussian $J$'s. Comparing the results to those
for binomial $J$'s \cite{dean} we see that $\langle N_s \rangle$
significantly depends on the distribution of $J$'s. It would be
quite interesting to investigate the dependence on the distribution
of $J$'s in a systematic way by calculating $\langle N_s \rangle$ 
for some other continuous distributions of $J$'s. One can try to do
this by applying the Tanaka-Edwards idea to distributions
of the type $p(J) = \int da f(a) e^{-J^2/a}$.
Another very challenging problem is to
calculate higher moments $\langle N_s^k \rangle$ 
and eventually also $\langle \ln N_s \rangle$.

\section*{Acknowledgments}
We thank W. Janke, A. Krzywicki and O. Martin for discussion.
We thank EC-RTN Network ``ENRAGE'' under grant No. MRTN-CT-2004-005616
and the Alexander von Humboldt Foundation for support.
Z.~B. acknowledges the support from a Marie Curie Actions Transfer of Knowledge project ``COCOS'',
Grant No. MTKD-CT-2004-517186 and a Polish Ministry of Science
and Information Society Technologies Grant 1P03B-04029 (2005-2008).

\section*{Appendix A}
In this appendix we shall prove that the matrix $\G$ is positive definite for 
$|g|<1$. We must show that 
\bq
	\x^T\G\x = \sum_{i,j} x_i G_{ij} x_j, \label{xgx}
\eq
is strictly positive for any vector $\x$ with non-zero length. From Eq. (\ref{gij}) we have:
\bq
	\x^T\G\x =  \sum_i \left( x_i^2 + g\frac{x_i}{\sqrt{k_i}} \sum_j A_{ij} \frac{x_j}{\sqrt{k_j}}\right). \label{xgx2}
\eq
Introducing new variables $y_i$: $x_i\equiv \sqrt{2k_i}y_i$
we can rewrite the right-hand side as:
\ba
	2(1-g)\sum_i k_i y_i^2 + g\sum_{i,j} A_{ij} (y_i+y_j)^2 \\
	=  2(1+g)\sum_i k_i y_i^2 + g\sum_{i,j} A_{ij} (y_i-y_j)^2. \label{sec}
\ea
The formula in the upper line shows that the quadratic form (\ref{xgx}) 
is non-negative for any $\x$ and for $0<g\leq 1$. It is zero only if $g=1$ and $\x=0$.
Therefore, $F(g)$ tends to a constant for $g\to 1$.
On the other hand, the second equation (\ref{sec}) tells us that for $-1<g<0$ it is positive definite as well.
However, for $g=-1$ we see that a zero mode appears. 
Because of the zero mode the Gaussian integration in Eq.~(\ref{start}) is divergent and consequently 
$F(g)\to\infty$ for $g\to -1$. 
These two observations indicate that the radius of convergence of $F(g)$ 
given by Eq.~(\ref{F_g}) is one, and that $f_l$ has the form:
\bq
	f_l = a_l (-1)^l l^{-\alpha},
\eq
where $\alpha\in(0,1]$ and $a_l$ tends to some constant for $l\to\infty$.

\section*{Appendix B}
\begin{figure}
\vspace{5mm}
\psfrag{xx}{$n$}\psfrag{yy}{$d_n$}
\includegraphics*[width=7cm]{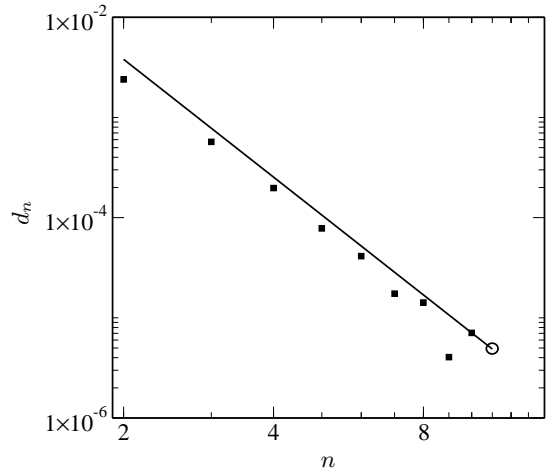}
\caption{\label{appb}Illustration of the procedure of estimating errors. 
We use as an example the data
for the SK model, for $L=11$ known coefficients $f_l$. 
We plot $(n,d_n)$ (squares) for $n=2,\dots,10$ in a log-log plot, and then fit a straight line. 
The line is shifted to be above all the data points. We extrapolate the line to obtain 
the value for $n=L=11$ (circle) and take this as an uncertainty of $F_\infty(1)$.}
\end{figure}

The estimation of systematic errors like those involved in the calculation of $F_\infty(1)$ 
from Eq. (\ref{estim}) is not an easy task. We believe, that the method described below gives 
an upper bound on the error of $F_\infty(1)$. Let us slightly adjust the notation for the sake
of clarity of the discussion. The value of $F_\infty(1)$ from the left-hand side of 
Eq. (\ref{estim}) depends on $L$, for which it has been estimated, 
so it is convenient to keep trace of $L$ in the notation.
We will denote the value of the estimate by $F^{(L)}_\infty(1)$. Of course it not the same
as $F_L(1)$ which is on the right-hand side of Eq. (\ref{estim}) and which means just a 
sum of the first $L$ coefficients $f_l$. The method relies on the following observation. 
We want to compute the deviation $D_L=F^{(L)}_\infty(1)-F_\infty(1)$ but we do not
know the limiting value $F_\infty(1)$. 
We can however compute slightly modified differences $d_n = F^{(n)}_\infty(1) - F^{(L)}_\infty(1)$ for all $n=2,\ldots,L-1$, and say that $D_n\approx d_n$ for $L$ being sufficiently large.
Now we can plot $d_n$ versus $n$ and extrapolate it to $n=L$ to obtain an estimator of $d_L$ which in turns estimates $D_L$ giving the error. 
Because $d_n$ falls with $n$ as a power of $n$ or faster, we can estimate $d_L$ from above by
fitting a straight line to the points $(\ln d_n,\ln n)$ and shifting it so that all points lie below it.
By extrapolating it to $n=L$ we get the value $d_L$ and use it to estimate the upper bound 
for the deviation between $F^{(L)}_\infty(1)$ and the true value $F_\infty(1)$. 
This procedure is illustrated in Fig. \ref{appb} for the case of SK model from Sec. III.
How reliable is this method? We checked that intervals 
$F(1)\pm \Delta F(1)$ obtained from $n=2,3,4,\dots,L-1$ first coefficients 
$f_l$ always include the value for $n=L$ for all graphs discussed in this paper. 
We checked also that reducing the error, say, by a factor two would result 
in many situations for which $F_L(1)$ would lie outside the error bars. This means that 
the method does not overestimate the error too much.

\end{document}